\documentclass[10pt,letterpaper,journal]{IEEEtran}
\usepackage{amsmath,amsfonts}
\usepackage{amssymb}
\usepackage{array}
\usepackage{tabularx}
\usepackage[caption=false,font=normalsize,labelfont=sf,textfont=sf]{subfig}
\usepackage{textcomp}
\usepackage{stfloats}
\usepackage{url}
\usepackage{verbatim}
\usepackage{graphicx}
\usepackage{cite}
\usepackage{booktabs}
\usepackage{multirow}
\usepackage[table]{xcolor}

\usepackage{algorithm}
\usepackage{algorithmic}

\usepackage{pifont}
\usepackage{balance}
\newcommand{\xmark}{\ding{55}}
\newcommand{\cmark}{\ding{51}}

\usepackage[hidelinks]{hyperref}

\begin{document}

\title{DF-PALM: Dynamic Fusion Prompt Learning for Audio--Language Models}

\author{Yulin Sun, Shenghua Li, Yong Dou, Xinwang Liu,~\IEEEmembership{Senior Member,~IEEE,} Cheng Yang,
and Kele Xu,~\IEEEmembership{Senior Member,~IEEE}%
\thanks{All authors are with the College of Computer Science and Technology,
National University of Defense Technology, Changsha, China.
Yulin Sun, Yong Dou, Cheng Yang, and Kele Xu are also with the State Key Laboratory of
Complex \& Critical Software Environment
(e-mail: sunyulin@nudt.edu.cn; lishenghua23@nudt.edu.cn;
yongdou@nudt.edu.cn; xinwangliu@nudt.edu.cn;
delpiero710@126.com; kele.xu@ieee.org).}%
\thanks{Corresponding author: Kele Xu.}}

\maketitle

\begin{abstract}
Few-shot audio classification requires audio--language models to distinguish variable acoustic events from limited labeled data. PALM's main formulation adapts encoded class representations without changing pretrained encoders, but its class representations remain input-independent and classification uses audio features without class-conditioned refinement. We introduce \textbf{DF-PALM}, a dynamic feature-space prompting framework that combines audio-conditioned class representations with asymmetric cross-attention for audio refinement. Class representations are conditioned after text encoding, held fixed across fusion layers for each input, and reused for final scoring. Both pretrained encoders remain frozen, avoiding audio-conditioned text-encoder passes. With PENGI as the frozen backbone, the main 16-shot evaluation covered 11 benchmarks and six baselines. DF-PALM achieved 80.69\% mean accuracy and 80.32\% macro-averaged F1, exceeding PALM by 4.11 and 4.30 percentage points, respectively. Paired benchmark-level tests supported both gains after Holm correction. Across four shot settings, DF-PALM achieved the highest mean accuracy and macro-averaged F1 among the four evaluated prompt-learning methods. The gains extended to LAION-CLAP and OpenFLAM, with significant improvements over PALM in both metrics. At 50 candidate classes, operator-counted inference floating-point operations increased by 0.39\% relative to PALM. These results support instance-conditioned feature-space adaptation as an effective approach to few-shot audio classification.
Our code is publicly available at GitHub \url{https://github.com/sunyulin0421/DF-PALM}.
\end{abstract}

\begin{IEEEkeywords}
Audio--Language Models, Few-Shot Learning, Prompt Learning, Audio Classification, Cross-Modal Fusion.
\end{IEEEkeywords}

\section{Introduction}

\begin{figure}[!t]
    \centering
    \includegraphics[width=\linewidth]{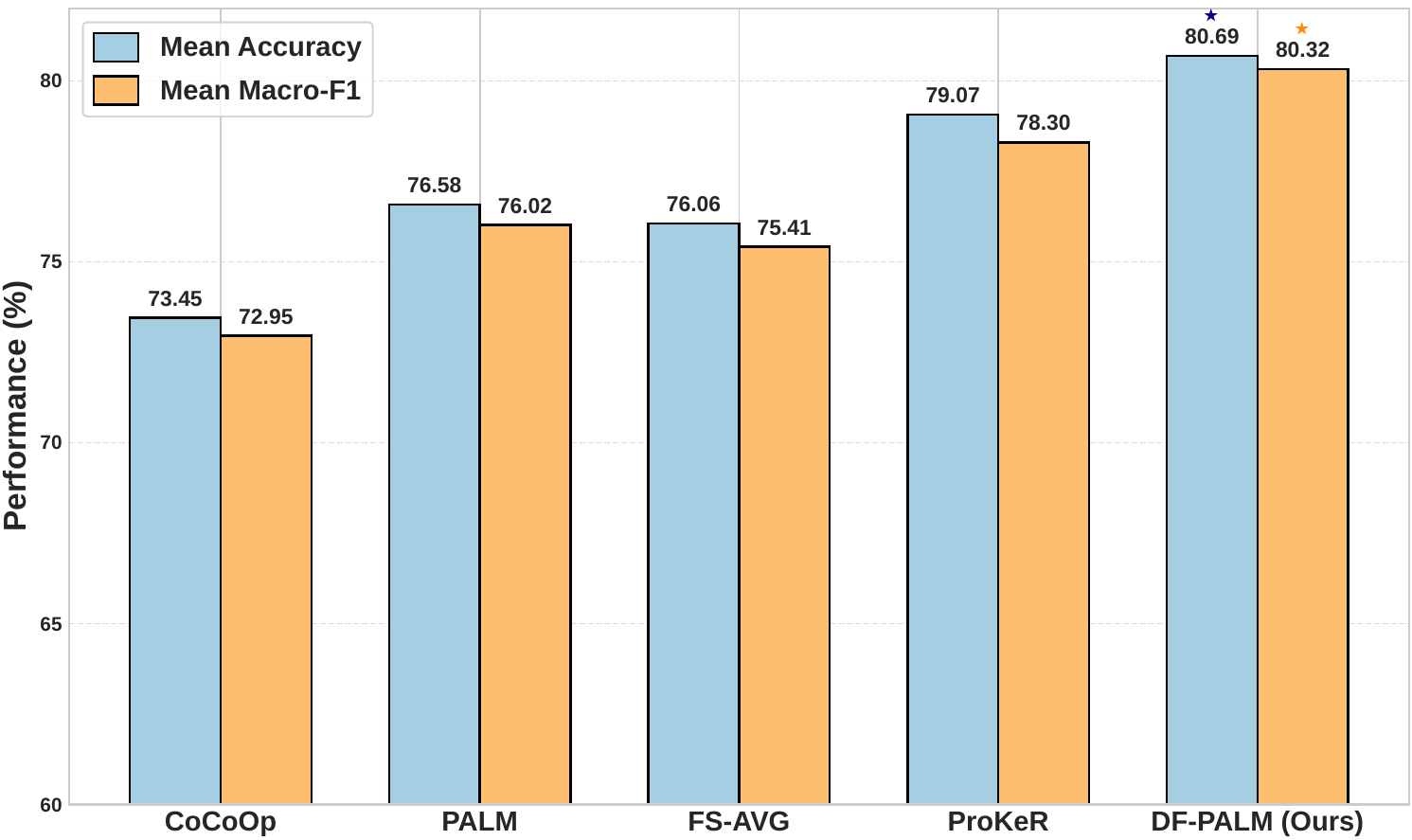}
    \caption{Cross-benchmark mean accuracy and Macro-F1 (\%) of the five methods shown across 11 audio classification benchmarks in the main 16-shot setting. Stars mark the highest mean for each metric among the methods shown.}
    \label{fig:performance}
\end{figure}

\IEEEPARstart{C}{onnecting} acoustic events with natural language is central to multimodal understanding~\cite{zhu2023languagebind,Wav2CLIP,tang2024salmonn}. Vision--Language Models have established strong correspondences between visual observations and semantic concepts~\cite{clip,jia2021align,alayrac2022flamingo}. Audio--Language Models (ALMs), by contrast, must align language with signals that unfold over time and vary markedly across recordings. Recordings of the same event can differ in source distance, background interference, and recording conditions~\cite{CLAP,audioclip,gong2021ast,fonseca2021fsd50k}. This variability makes reliable audio--language alignment particularly difficult when only a few labeled examples are available.

Prompt learning adapts ALMs without fine-tuning the full backbone, using input-independent contexts or instance-conditioned prompts formed before text encoding~\cite{hanif2024palm,coop,zhou2022cocoop}. CoOp and CoCoOp optimize input prompts by propagating gradients through the text encoder while retaining its pretrained weights. The main PALM formulation instead adapts encoded text features with input-independent class representations. Acoustic variation motivates conditioning these representations on each input while retaining class-specific information. These prompt baselines classify audio features by similarity to class representations, without using class information to refine the audio features before scoring~\cite{hanif2024palm}.

Pre-scoring audio--text interaction has been explored through parameter-free alignment and multi-grained correspondence learning~\cite{seth2025pat,zhao2024multigrained}. The question here is how to combine instance-conditioned class representations with audio refinement after text encoding while keeping pretrained encoders fixed.

We present \textbf{DF-PALM} (\textbf{D}ynamic \textbf{F}usion \textbf{P}rompt Learning for \textbf{A}udio--\textbf{L}anguage \textbf{M}odels), a dynamic feature-space prompt-learning framework for frozen ALMs. DF-PALM combines an \textbf{Instance-Level Dynamic Context Generator} with \textbf{Asymmetric Cross-Modal Fusion}. The first adds a shared audio-conditioned residual to class-specific contexts after text encoding. The second refines the audio representation against the conditioned class representations, which remain fixed across fusion layers for each input. This design combines instance conditioning with class-guided audio refinement in feature space, without requiring audio-conditioned text-encoder passes.

Our main contributions are as follows:
\begin{itemize}
    \item DF-PALM couples post-encoding instance conditioning with class-guided audio refinement through a conditioned class matrix shared by fusion and final scoring, while keeping both pretrained encoders fixed.
    
    \item Across 11 benchmarks in the main 16-shot setting, DF-PALM achieved the highest cross-benchmark mean accuracy and Macro-F1 among the evaluated methods (Appendix~\ref{app:main_f1_results}; Fig.~\ref{fig:performance}). It led the four evaluated prompt-learning methods on both metrics across all four shot settings. Capacity controls and evaluations on two additional backbones further supported the proposed feature-space design.
    
    \item We quantify the computational cost of DF-PALM at both module and end-to-end levels, separating operator-counted FLOPs from measured latency and memory.
\end{itemize}

\section{Related Work}

\subsection{Audio--Language Representation Learning}

Audio--Language Models (ALMs) learn shared representations that connect acoustic signals with natural-language descriptions~\cite{huang2022mulan,mei2023wavcaps}. Contrastive models such as AudioCLIP~\cite{audioclip} and CLAP~\cite{CLAP,laionclap2023} align global audio and text embeddings for zero-shot transfer. Generative ALMs such as PENGI~\cite{deshmukh2023pengi} connect audio encoders to pretrained language models through prefixes, adapters, or related interfaces~\cite{Qwen-Audio,gong2023listen}.

Despite their broad transfer capabilities, adapting ALMs to data-scarce downstream tasks can still require substantial optimization. This has motivated approaches that adapt pretrained representations without updating the full backbone~\cite{houlsby2019parameter,zaken2021bitfit}.

\subsection{Prompt Learning in Multimodal Models}

Prompt learning adapts pretrained models through learnable context parameters rather than full-backbone updates~\cite{prefix-tuning,Liu2023}. Initially developed for natural-language models~\cite{brown2020language,lester2021power}, this approach was subsequently extended to multimodal architectures.

In vision--language models, CoOp~\cite{coop} replaces hand-crafted prompts with learnable but input-independent context vectors. CoCoOp~\cite{zhou2022cocoop} introduces instance conditioning by generating context vectors from visual features. Because CoCoOp's conditioned prompts are formed before text encoding, their optimization requires gradients to pass through the text encoder. Other developments include multi-modal prompt learning~\cite{khattak2023maple} and prompt distribution learning~\cite{lu2022prompt}.

PALM~\cite{hanif2024palm} brings prompt learning to ALMs by optimizing encoded text features rather than input tokens. This design moves learnable adaptation into feature space and leaves the text-encoder parameters unchanged, with input-independent class representations in its main formulation. PALM's ablations also explore audio feedback to encoded text features. DF-PALM builds on this feature-space setting by coupling instance-conditioned class representations with audio refinement against a class matrix held fixed across fusion layers and reused for final scoring.

\subsection{Cross-Modal Alignment and Parameter-Efficient Adaptation}

Fusion encoders learn interactions between modalities before prediction~\cite{ALBEF,BLIP2,kim2021vilt,yu2022coca}, and audio--language methods also introduce alignment after pretraining. PAT selects and weights textual prompts and uses text-guided frame-level audio reweighting to improve zero-shot classification without additional trainable parameters~\cite{seth2025pat}. Zhao et al. learn multi-grained audio--text correspondences using a key-value cache, cross-modal contrast, and optimal transport for few-shot audio recognition~\cite{zhao2024multigrained}. DF-PALM learns instance conditioning and asymmetric fusion directly from global audio vectors and encoded class names, retaining both pretrained encoders.

Parameter-efficient adaptation also includes inserted adapters~\cite{houlsby2019parameter} and low-rank adaptation (LoRA), which learns low-rank updates to pretrained weight matrices~\cite{hu2022lora}. The three-stage ASPL$^{*}$ variant introduces affine modulation within the audio encoder and combines it with text-side prompt learning~\cite{cho2026aspl}. These approaches modify encoder processing through inserted modules or weight updates. DF-PALM places its trainable conditioning and fusion modules after the pretrained encoders, operating on their output representations.

Support-set adaptation provides a further approach to few-shot classification using frozen representations. ProKeR combines a zero-shot predictor with support information through kernel adaptation~\cite{bendou2025proker}, whereas FS-AVG averages support audio embeddings into class prototypes~\cite{taylor25_interspeech}. We evaluate both methods with the same PENGI backbone used by DF-PALM.

% Table~\ref{table:comparison} describes optimization space, context type, and training time for the prompt-learning methods in the ESC-50 8-shot setting. Static and dynamic contexts describe whether the class representation depends on the input.
For the selected configurations in the matched ESC-50 8-shot setting, DF-PALM required 1 h 27 min of training, compared with 5 h 51 min for CoCoOp and 1 h 24 min for PALM (Table~\ref{table:comparison}).

\begin{table}[!t]
\centering
\setlength{\tabcolsep}{1mm}
\caption{Comparison of Prompt-Learning Paradigms. Learned indicates whether prompt context is optimized; Opt.\ Space denotes the optimization space. Static/Dynamic indicate whether context depends on the current audio input. Times were measured in the matched ESC-50 8-shot setting.}
\label{table:comparison}
\begin{tabular}{lcccc}
\toprule
\textbf{Method} & \textbf{Learned} & \textbf{Opt.\ Space} & \textbf{Context} & \textbf{Time} \\
\midrule
Zero-Shot & \xmark & N/A     & Manual  & N/A \\
CoOp      & \cmark & Input   & Static  & 1~h 28~min \\
CoCoOp    & \cmark & Input   & Dynamic & 5~h 51~min \\
PALM      & \cmark & Feature & Static  & 1~h 24~min \\
\midrule
\textbf{DF-PALM} & \cmark & \textbf{Feature} & \textbf{Dynamic} & \textbf{1~h 27~min} \\
\bottomrule
\end{tabular}
\end{table}

\begin{figure*}[!t]
    \centering
    \includegraphics[width=0.98\textwidth]{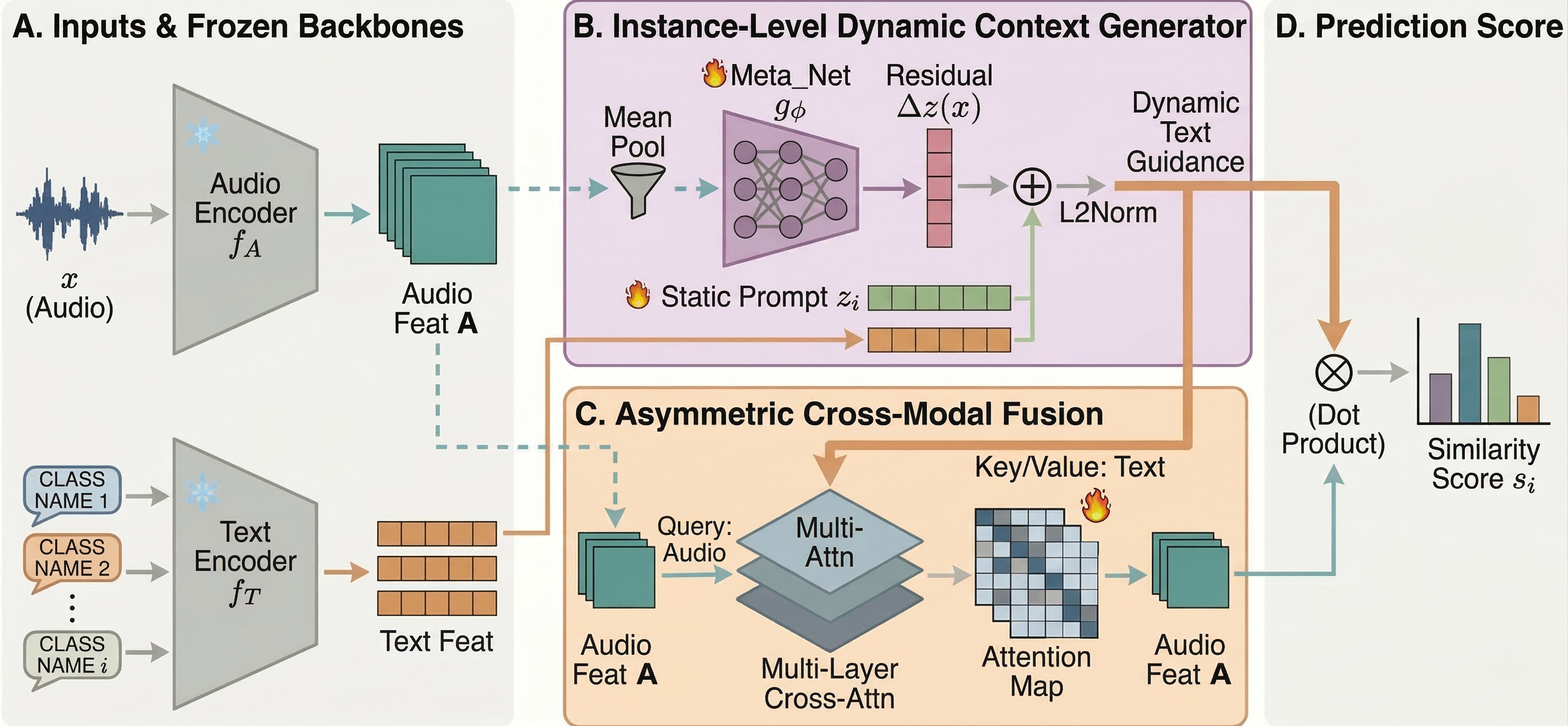}
    \caption{Overview of DF-PALM. (A) Frozen encoders produce global audio and base class representations; Mean Pool denotes the backbone's aggregation into a global audio vector. (B) Meta\_Net denotes the dynamic context generator \(g_{\phi}\), and Dynamic Text Guidance denotes the conditioned class matrix \(\mathbf{T}(x)\). Class-specific mixing follows Equation~\eqref{eq:dynamic_context}. (C) Asymmetric fusion refines the audio representation against this matrix. (D) Normalized similarity scoring uses the logit scale in Equation~\eqref{eq:prediction}.}
    \label{fig:framework}
\end{figure*}

\section{Methodology}
\label{sec:method}

DF-PALM first conditions encoded class representations on each audio input and then uses them to refine the corresponding audio representation (Fig.~\ref{fig:framework}). All trainable adaptation operates in feature space, while the pretrained audio and text encoders remain fixed.

\subsection{Problem Formulation}

Let \(\mathcal{D}=\{(x_n,y_n)\}_{n=1}^{N}\) denote a labeled audio dataset, where \(x_n\) is an audio waveform and \(y_n\in\{1,\ldots,C\}\) is its class label. We instantiate DF-PALM with the pretrained PENGI audio--language backbone, comprising a frozen audio encoder \(f_A\) and a frozen text encoder \(f_T\)~\cite{deshmukh2023pengi}.

For an input \(x\), the audio encoder produces a global representation \(\mathbf{a}\in\mathbb{R}^{d}\). The encoder includes the backbone's pooling and projection, so the context generator receives one global vector per waveform. For class \(i\), the fixed input \(p_i\) is its unmodified class-name string, specified in Appendix~\ref{app:datasets}. Its base class representation is \(\mathbf{t}_i\in\mathbb{R}^{d}\):
\begin{equation}
    \mathbf{a}=\operatorname{L2Norm}\!\left(f_A(x)\right),\quad
    \mathbf{t}_i=\operatorname{L2Norm}\!\left(f_T(p_i)\right).
\end{equation}
For each forward batch, the frozen text encoder computes the base class representations once and shares them across samples in the batch.

Following PALM~\cite{hanif2024palm}, feature-space prompt learning associates each class with a trainable context vector \(\mathbf{z}_i\in\mathbb{R}^{d}\). A learnable scalar \(\beta_i\) determines the class-specific mixing coefficient \(\lambda_i=\operatorname{sigmoid}(\beta_i)\). Static feature-space prompting forms
\begin{equation}
    \widetilde{\mathbf{t}}_i
    =
    \operatorname{L2Norm}\!\left(
    (1-\lambda_i)\mathbf{t}_i+\lambda_i\mathbf{z}_i
    \right).
    \label{eq:static_prompt}
\end{equation}
Both \(\mathbf{z}_i\) and \(\lambda_i\), however, remain independent of the audio input.

\subsection{Instance-Level Dynamic Context Generation}

The Instance-Level Dynamic Context Generator conditions the feature-space context on the global audio representation. A bottleneck network \(g_{\phi}\) predicts an input-dependent residual:
\begin{equation}
    \Delta\mathbf{z}(x)
    =
    g_{\phi}(\mathbf{a})
    =
    \mathbf{W}_{2}
    \rho\!\left(\mathbf{W}_{1}\mathbf{a}\right),
    \label{eq:dynamic_residual}
\end{equation}
where \(\rho\) denotes ReLU. We use \(\mathbf{W}_{1}\in\mathbb{R}^{\frac{d}{r}\times d}\) and \(\mathbf{W}_{2}\in\mathbb{R}^{d\times\frac{d}{r}}\), with a reduction ratio of \(r=16\). The bottleneck reduces the hidden width from \(d\) to \(d/r\).

The residual is shared across classes and added to each class-specific context vector:
\begin{equation}
\begin{aligned}
    &\mathbf{z}_i(x)=\mathbf{z}_i+\Delta\mathbf{z}(x),\\
    &\mathbf{T}_i(x)=\operatorname{L2Norm}\!\left((1-\lambda_i)\mathbf{t}_i+\lambda_i\mathbf{z}_i(x)\right).
\end{aligned}
\label{eq:dynamic_context}
\end{equation}
Each conditioned class representation combines the shared audio-dependent residual \(\Delta\mathbf{z}(x)\) with its class-specific context vector \(\mathbf{z}_i\) and mixing coefficient \(\lambda_i\). Class-specific mixing and L2 normalization determine how this residual changes each class representation. The shared context generator provides instance conditioning with a parameter count independent of the number of classes. Conditioning follows text encoding and requires no audio-conditioned text-encoder pass; the text-encoder parameters remain frozen.

\subsection{Asymmetric Cross-Modal Fusion}
\label{sec:fusion}

The instance-conditioned class representations are stacked as
\begin{equation}
    \mathbf{T}(x)
    =
    [\mathbf{T}_1(x);\ldots;\mathbf{T}_C(x)]
    \in\mathbb{R}^{C\times d}.
    \label{eq:class_matrix}
\end{equation}
Let \(\mathbf{h}^{(0)}=\mathbf{a}\). At fusion layer \(l\), the current audio state forms a single query token, while \(\mathbf{T}(x)\) provides the keys and values:
\begin{equation}
    \mathbf{m}^{(l)}
    =
    \operatorname{MHA}_{l}\!\left(
    \mathbf{h}^{(l-1)},
    \mathbf{T}(x),
    \mathbf{T}(x)
    \right),
    \qquad l=1,\ldots,L,
    \label{eq:cross_attention}
\end{equation}
where \(\operatorname{MHA}_{l}\) denotes multi-head cross-attention~\cite{vaswani2017attention}. A post-norm residual update gives
\begin{align}
    \widetilde{\mathbf{h}}^{(l)}
    &=
    \mathbf{h}^{(l-1)}
    +
    \operatorname{Dropout}\!\left(\mathbf{m}^{(l)}\right),\\
    \mathbf{h}^{(l)}
    &=
    \operatorname{L2Norm}\!\left(
    \operatorname{LayerNorm}
    \left(\widetilde{\mathbf{h}}^{(l)}\right)
    \right).
    \label{eq:fusion_update}
\end{align}
For each input, the conditioned class matrix \(\mathbf{T}(x)\) remains fixed across fusion layers while the audio state \(\mathbf{h}^{(l)}\) is refined. Each layer uses this matrix as its key and value input and updates only the audio state. This direction of updating defines the asymmetric structure. During training, the classification loss optimizes the context and fusion parameters through this computation.

\subsection{Prediction and Optimization}

For sample \(x_n\), let \(\mathbf{T}_{n,i}\equiv\mathbf{T}_i(x_n)\) denote the conditioned representation of class \(i\), and let \(\mathbf{h}_n^{(L)}\) denote the corresponding refined audio representation. The class logit is
\begin{equation}
    s_{n,i}
    \equiv
    s_i(x_n)
    =
    \tau
    \left\langle
    \mathbf{h}_n^{(L)},\mathbf{T}_{n,i}
    \right\rangle,
    \label{eq:prediction}
\end{equation}
where \(\langle\cdot,\cdot\rangle\) denotes the inner product and \(\tau=100\) is the fixed logit scale used in the implementation. Because both representations are L2-normalized, the inner product is equivalent to cosine similarity.

For a mini-batch of size \(B\), the cross-entropy objective is
\begin{equation}
    \mathcal{L}
    =
    -\frac{1}{B}
    \sum_{n=1}^{B}
    \log
    \frac{\exp(s_{n,y_n})}
    {\sum_{i=1}^{C}\exp(s_{n,i})}.
    \label{eq:training_loss}
\end{equation}
Training optimizes the class-specific prompt parameters \(\{\mathbf{z}_i,\beta_i\}\), the context generator \(g_{\phi}\), and the cross-attention layers, while the pretrained audio and text encoders remain frozen. Algorithm~\ref{alg:dfpalm} summarizes the batch forward pass; the dynamic class representations are constructed separately for each audio input.

\begin{algorithm}[!t]
\caption{Batch forward pass of DF-PALM}
\label{alg:dfpalm}
\textbf{Input}: Audio mini-batch
\(\mathcal{X}=\{x_n\}_{n=1}^{B}\);
fixed class-prompt batch
\(\mathcal{P}=\{p_i\}_{i=1}^{C}\);
frozen encoders \(f_A,f_T\)\\
\textbf{Parameter}: Context generator \(g_{\phi}\);
prompt parameters \(\{\mathbf{z}_i,\beta_i\}_{i=1}^{C}\);
\(L\) fusion layers; fixed logit scale \(\tau=100\)\\
\textbf{Output}: Logit matrix
\(\mathbf{S}=[s_{n,i}]\in\mathbb{R}^{B\times C}\)

\begin{algorithmic}[1]
\STATE \textit{Compute representations shared across the mini-batch}
\STATE \(\mathbf{T}_{\mathrm{base}}
\gets f_T(\mathcal{P})\)

\FOR{\(i=1,\ldots,C\)}
    \STATE \(\mathbf{t}_i\gets
    \operatorname{L2Norm}
    \!\left(\mathbf{T}_{\mathrm{base}}[i,:]\right)\)
    \STATE \(\lambda_i\gets
    \operatorname{sigmoid}(\beta_i)\)
\ENDFOR

\STATE \(\mathbf{A}\gets f_A(\mathcal{X})\)

\FOR{\(n=1,\ldots,B\)}
    \STATE \(\mathbf{a}_n\gets
    \operatorname{L2Norm}
    \!\left(\mathbf{A}[n,:]\right)\)
    \STATE \(\Delta\mathbf{z}_n
    \gets g_{\phi}(\mathbf{a}_n)\)

    \FOR{\(i=1,\ldots,C\)}
        \STATE \(\mathbf{u}_{n,i}\gets
        (1-\lambda_i)\mathbf{t}_i+
        \lambda_i
        \left(\mathbf{z}_i+\Delta\mathbf{z}_n\right)\)
        \STATE \(\mathbf{T}_{n,i}\gets
        \operatorname{L2Norm}
        \!\left(\mathbf{u}_{n,i}\right)\)
    \ENDFOR

    \STATE \(\mathbf{T}_n\gets
    [\mathbf{T}_{n,1};\ldots;\mathbf{T}_{n,C}]\)
    \STATE \(\mathbf{h}_n^{(0)}\gets\mathbf{a}_n\)

    \FOR{\(l=1,\ldots,L\)}
        \STATE \(\mathbf{m}_n^{(l)}\gets
        \operatorname{MHA}_{l}
        \!\left(
        \mathbf{h}_n^{(l-1)},
        \mathbf{T}_n,
        \mathbf{T}_n
        \right)\)
        \STATE \(\widetilde{\mathbf{h}}_n^{(l)}
        \gets
        \mathbf{h}_n^{(l-1)}
        +
        \operatorname{Dropout}
        \!\left(\mathbf{m}_n^{(l)}\right)\)
        \STATE \(\mathbf{h}_n^{(l)}
        \gets
        \operatorname{L2Norm}
        \!\left(
        \operatorname{LayerNorm}
        \!\left(\widetilde{\mathbf{h}}_n^{(l)}\right)
        \right)\)
    \ENDFOR

    \FOR{\(i=1,\ldots,C\)}
        \STATE \(s_{n,i}\gets
        \tau
        \left\langle
        \mathbf{h}_n^{(L)},
        \mathbf{T}_{n,i}
        \right\rangle\)
    \ENDFOR
\ENDFOR

\STATE \textbf{return} \(\mathbf{S}\)
\end{algorithmic}
\end{algorithm}

\subsection{Design Rationale and Complexity}
\label{sec:complexity}

\paragraph{Asymmetric update}
Instance conditioning first adapts the class representations to the input. Asymmetric fusion then refines the audio state against the same class matrix used for final scoring, separating class conditioning from audio refinement. Section~\ref{sec:ablation} evaluates the contribution of audio refinement and compares this design with joint self-attention.

\paragraph{Computational complexity}
We characterize the incremental computational cost of the adaptation modules beyond the frozen backbone. The context generator requires \(\mathcal{O}(d^2/r)\) operations per input, followed by \(\mathcal{O}(Cd)\) operations to construct the instance-conditioned class representations. At each fusion layer, attention scoring and value aggregation require \(\mathcal{O}(Cd)\) operations. Including the query, key, value, and output projections, the fusion cost is
\begin{equation}
    \mathcal{O}\!\left(
    L\left[(C+1)d^2+Cd\right]
    \right).
    \label{eq:complexity}
\end{equation}
For fixed \(d\) and \(L\), this cost grows linearly with the number of classes.
The \(C\)-dependent terms correspond to constructing and attending over the class matrix, which is the behavior isolated by the module-level scalability profile in Section~\ref{sec:scalability}.

\begin{table*}[!t]
\centering
\caption{Complete Accuracy Results (\%) Across 11 Audio Classification Benchmarks in the Main 16-Shot Setting. Zero-Shot uses no support samples; FS-AVG and ProKeR require no gradient-based training. S-0--S-2 denote matched seeds, including support subsets, and AVG denotes their mean. Folded benchmarks are averaged within each seed; AVERAGE is the unweighted cross-benchmark mean. Bold marks the highest mean across all seven methods in each row.}
\label{table:main_results}
{\footnotesize
\renewcommand{\arraystretch}{1.08}
\setlength{\tabcolsep}{1.5pt}
\begin{tabular*}{\textwidth}{@{\extracolsep{\fill}}l c cccc cccc@{}}
\toprule
\multicolumn{10}{l}{\textbf{Panel A: Zero-Shot and Input-Space Prompt Learning}} \\
\midrule
\multirow{2}{*}{\textbf{Benchmark}} & \textbf{Zero-Shot} & \multicolumn{4}{c}{\textbf{CoOp}} & \multicolumn{4}{c}{\textbf{CoCoOp}} \\
\cmidrule(lr){2-2}
\cmidrule(lr){3-6}
\cmidrule(lr){7-10}
 & -- & S-0 & S-1 & S-2 & \textbf{AVG} & S-0 & S-1 & S-2 & \textbf{AVG} \\
\midrule
Beijing-Opera & 28.81 & 93.23 & 96.60 & 96.19 & 95.34 & 95.77 & 98.30 & 99.16 & 97.74 \\
CREMA-D & 23.10 & 31.30 & 41.97 & 27.60 & 33.62 & 25.39 & 33.58 & 31.56 & 30.18 \\
ESC50-Actions & 65.25 & 96.25 & 94.00 & 95.50 & 95.25 & 96.31 & 96.20 & 96.48 & 96.33 \\
ESC-50 & 49.65 & 94.10 & 93.90 & 93.45 & 93.82 & 94.60 & 93.70 & 94.50 & 94.27 \\
GT-Music-Genre & 32.50 & 72.50 & 69.50 & 73.50 & 71.83 & 75.00 & 74.50 & 76.00 & 75.17 \\
NS-Instruments & 32.91 & 57.28 & 55.62 & 61.77 & 58.22 & 59.96 & 57.40 & 64.38 & 60.58 \\
RAVDESS & 12.22 & 38.49 & 26.88 & 34.22 & 33.20 & 37.27 & 43.99 & 35.23 & 38.83 \\
SESA & 72.38 & 91.43 & 89.52 & 87.62 & 89.52 & 83.81 & 87.62 & 89.52 & 86.98 \\
TUT2017 & 24.35 & 63.91 & 66.67 & 65.25 & 65.28 & 74.99 & 72.15 & 73.12 & 73.42 \\
UrbanSound8K & 53.49 & 76.00 & 73.78 & 76.66 & 75.48 & 75.76 & 77.84 & 75.97 & 76.52 \\
VocalSound & 41.97 & 71.62 & 74.85 & 66.42 & 70.96 & 80.81 & 78.25 & 74.63 & 77.90 \\
\midrule
\textbf{AVERAGE} & 39.69 & 71.46 & 71.21 & 70.74 & 71.14 & 72.70 & 73.96 & 73.69 & 73.45 \\
\bottomrule
\end{tabular*}
\par\vspace{7pt}
\begin{tabular*}{\textwidth}{@{\extracolsep{\fill}}l cccc cccc cccc cccc@{}}
\toprule
\multicolumn{17}{l}{\textbf{Panel B: Support-Set and Feature-Space Adaptation}} \\
\midrule
\multirow{2}{*}{\textbf{Benchmark}} & \multicolumn{4}{c}{\textbf{FS-AVG (PENGI)}} & \multicolumn{4}{c}{\textbf{ProKeR (PENGI)}} & \multicolumn{4}{c}{\textbf{PALM}} & \multicolumn{4}{c}{\textbf{DF-PALM (Ours)}} \\
\cmidrule(lr){2-5}
\cmidrule(lr){6-9}
\cmidrule(lr){10-13}
\cmidrule(lr){14-17}
 & S-0 & S-1 & S-2 & \textbf{AVG} & S-0 & S-1 & S-2 & \textbf{AVG} & S-0 & S-1 & S-2 & \textbf{AVG} & S-0 & S-1 & S-2 & \textbf{AVG} \\
\midrule
Beijing-Opera & 99.58 & 100.00 & 100.00 & \textbf{99.86} & 94.93 & 95.32 & 97.46 & 95.90 & 97.47 & 90.66 & 97.87 & 95.33 & 98.74 & 100.00 & 100.00 & 99.58 \\
CREMA-D & 38.62 & 44.66 & 26.73 & 36.67 & 40.16 & 35.66 & 39.29 & 38.37 & 44.53 & 35.80 & 23.44 & 34.59 & 44.06 & 43.05 & 39.15 & \textbf{42.09} \\
ESC50-Actions & 98.75 & 99.25 & 98.75 & 98.92 & 98.25 & 99.50 & 97.75 & 98.50 & 97.00 & 96.25 & 96.50 & 96.58 & 98.50 & 99.25 & 99.25 & \textbf{99.00} \\
ESC-50 & 96.90 & 97.05 & 97.10 & 97.02 & 97.15 & 96.85 & 96.70 & 96.90 & 95.60 & 96.00 & 96.20 & 95.93 & 97.40 & 96.95 & 97.20 & \textbf{97.18} \\
GT-Music-Genre & 78.50 & 80.50 & 78.50 & 79.17 & 79.50 & 78.00 & 78.50 & 78.67 & 79.00 & 78.50 & 82.50 & 80.00 & 79.50 & 81.50 & 82.50 & \textbf{81.17} \\
NS-Instruments & 64.55 & 64.72 & 66.89 & 65.39 & 69.46 & 74.27 & 73.78 & \textbf{72.50} & 63.94 & 61.08 & 66.48 & 63.83 & 67.94 & 72.63 & 73.95 & 71.51 \\
RAVDESS & 31.57 & 34.62 & 32.18 & 32.79 & 41.55 & 43.38 & 47.25 & 44.06 & 45.62 & 46.03 & 46.23 & 45.96 & 47.05 & 49.08 & 50.92 & \textbf{49.02} \\
SESA & 90.48 & 94.29 & 86.67 & 90.48 & 93.33 & 94.29 & 91.43 & 93.02 & 88.57 & 91.43 & 88.57 & 89.52 & 91.43 & 96.19 & 94.29 & \textbf{93.97} \\
TUT2017 & 68.62 & 68.08 & 67.08 & 67.93 & 80.04 & 81.37 & 80.79 & 80.73 & 79.59 & 80.47 & 77.29 & 79.12 & 84.37 & 84.82 & 83.93 & \textbf{84.37} \\
UrbanSound8K & 85.73 & 86.55 & 86.90 & \textbf{86.39} & 86.39 & 85.39 & 86.68 & 86.15 & 81.20 & 80.37 & 80.74 & 80.77 & 85.68 & 86.42 & 85.73 & 85.94 \\
VocalSound & 82.23 & 81.12 & 82.65 & 82.00 & 83.18 & 85.24 & 86.52 & \textbf{84.98} & 81.01 & 81.68 & 79.64 & 80.78 & 83.68 & 83.32 & 84.29 & 83.76 \\
\specialrule{\lightrulewidth}{0pt}{0pt}
\textbf{AVERAGE} & 75.96 & 77.35 & 74.86 & 76.06 & 78.54 & 79.02 & 79.65 & 79.07 & 77.59 & 76.21 & 75.95 & 76.58 & 79.85 & 81.20 & 81.02 & \textbf{80.69} \\
\bottomrule
\end{tabular*}
}
\end{table*}

\begin{table}[!t]
\centering
\small
\setlength{\tabcolsep}{2.5pt}
\caption{Paired 16-Shot Comparison Between DF-PALM and PALM Across 11 Benchmarks. Means weight benchmarks equally after fold and seed aggregation. \(\Delta\) (percentage points) and 95\% bootstrap percentile intervals use benchmark-level DF-PALM minus PALM differences. \(p_{\mathrm{adj}}\) denotes Holm-adjusted exact two-sided Wilcoxon signed-rank \(p\)-values.}
\label{table:statistical_comparison}
\begin{tabular}{lcccc}
\toprule
\textbf{Metric} & \textbf{PALM} & \textbf{DF-PALM} &
\(\boldsymbol{\Delta}\) \textbf{[95\% CI]} &
\(\boldsymbol{p_{\mathrm{adj}}}\) \\
\midrule
Accuracy & 76.58 & 80.69 &
\(+4.11\ [2.90,\,5.37]\) &
\(1.95\times10^{-3}\) \\
Macro-F1 & 76.02 & 80.32 &
\(+4.30\ [2.99,\,5.67]\) &
\(1.95\times10^{-3}\) \\
\bottomrule
\end{tabular}
\end{table}

\section{Experiments}
\label{sec:experiments}

The main evaluation used PENGI's pretrained encoders for sequence-level classification across 11 audio classification benchmarks. The autoregressive decoder was excluded from prediction and the reported parameter counts. The experiments assessed performance across acoustic tasks and shot settings, component contributions, configuration sensitivity, and computational cost. Comparisons covered prompt learning, support-set kernel adaptation (ProKeR), and training-free audio-prototype adaptation (FS-AVG). An additional 16-shot evaluation with LAION-CLAP and OpenFLAM assessed whether the mean gains extended across audio--language backbones.

\subsection{Datasets and Evaluation Metrics}

The 11 benchmarks covered acoustic scenes, sound events, speech emotion, musical instruments, and music genres, with 4--50 classes. We followed the PALM partition protocols with disjoint support and test samples. Benchmark specifications and split protocols are given in Appendix~\ref{app:datasets}.

We report classification accuracy and macro-averaged F1 score (Macro-F1)~\cite{sokolova2009systematic}. Macro-F1 is the unweighted mean of class-wise F1 scores and was computed identically for all methods. Scores were averaged across folds within each seed and then across the three matched seeds within each benchmark. Cross-benchmark aggregates weight all 11 benchmarks equally, and paired tests use one aggregated value per benchmark.

\subsection{Baselines}

We compared DF-PALM with six baselines spanning zero-shot inference, prompt learning, kernel adaptation, and prototype adaptation. For CoOp and CoCoOp, we replaced the image encoder with PENGI's audio encoder while retaining their prompt-learning procedures.

Zero-Shot~\cite{deshmukh2023pengi} uses the template ``\textit{This is a recording of \{CLASS\}}''. CoOp~\cite{coop} learns input-independent context tokens before text encoding, whereas CoCoOp~\cite{zhou2022cocoop} conditions them on each audio input. PALM~\cite{hanif2024palm} optimizes input-independent class representations after text encoding and is the principal feature-space baseline.

ProKeR (PENGI) applies the ProKeR kernel formulation~\cite{bendou2025proker} to frozen PENGI audio and class-text embeddings. FS-AVG (PENGI)~\cite{taylor25_interspeech} averages the 16-shot audio embeddings within each class and classifies by cosine similarity between normalized queries and prototypes, without text embeddings, gradients, or trainable parameters.

\subsection{Implementation Details}

The main evaluation used frozen PENGI encoders and 16 support examples per class. All methods used identical class-wise support subsets. DF-PALM selected its fusion configuration independently within each support set; ProKeR selected its kernel configuration within those supports. Full split and selection protocols are given in Appendices~\ref{app:datasets} and~\ref{app:implementation}.

All few-shot methods used seeds 0, 1, and 2. DF-PALM retained each run's 16-shot-selected configuration for the other shot settings. Optimization and implementation details are given in Appendix~\ref{app:implementation}.

\begin{figure*}[!t]
\centering
\includegraphics[width=0.95\linewidth]{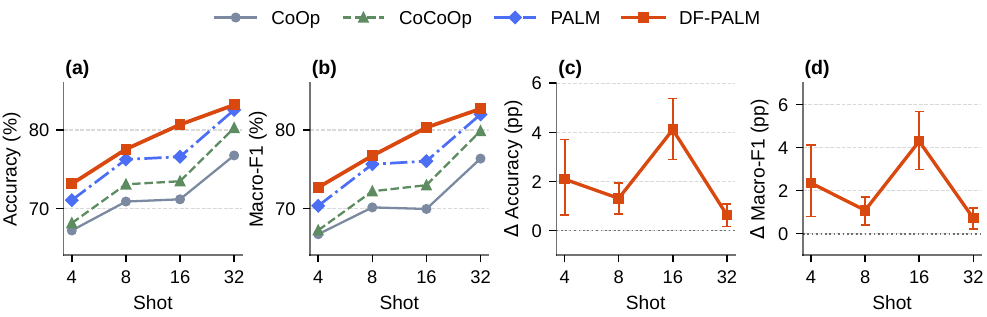}
\caption{Cross-benchmark performance across shot settings. Panels (a) and (b) show mean accuracy and Macro-F1 for four prompt-learning methods; panels (c) and (d) show paired DF-PALM-minus-PALM differences. Markers denote mean benchmark-level differences, error bars denote 95\% bootstrap percentile intervals from 200,000 resamples, and dotted lines mark zero. Scores use fold and seed aggregation with equal benchmark weighting; 16-shot points reproduce the main experiment.}
\label{fig:few_shot}
\end{figure*}

\begin{table*}[!t]
\centering
\footnotesize
\setlength{\tabcolsep}{3pt}
\caption{Capacity and Adaptation Controls in the 16-Shot Setting (\%). All methods use the same training settings. Values are unweighted means across 11 benchmarks.}
\label{table:capacity_controls}
\resizebox{0.90\textwidth}{!}{%
\begin{tabular}{@{}l l c c l rr@{}}
\toprule
\multicolumn{1}{@{}l}{\textbf{Method}} & \textbf{Adaptation space} & \textbf{Dynamic class} & \textbf{Cross-modal} & \textbf{Trainable parameters} & \textbf{Acc.} & \multicolumn{1}{r@{}}{\textbf{Macro-F1}} \\
 \multicolumn{1}{@{}l}{}& & \textbf{representation} & \textbf{fusion} & & & \multicolumn{1}{r@{}}{}\\
\midrule
PALM & Encoder output & No & No & Prompt only & 76.58 & 76.02 \\
Linear Probe & Frozen audio output & No & No & Linear head & 77.29 & 76.38 \\
Static-PALM+MLP & Encoder output & No & No & 12.60M audio MLP & 79.47 & 79.11 \\
PALM+Audio-LoRA & Audio encoder internal & No & No & 141,312 LoRA + prompt & 79.27 & 78.93 \\
PALM+ASPL$^{*}$ & Audio encoder internal & No & No & 704 ASPL + prompt & 75.57 & 75.08 \\
\multicolumn{1}{@{}l}{\textbf{DF-PALM}} & \textbf{Encoder output} & \textbf{Yes} & \textbf{Yes} & \textbf{12.78M} & \textbf{80.69} & \multicolumn{1}{r@{}}{\textbf{80.32}} \\
\bottomrule
\end{tabular}%
}
\end{table*}

\subsection{Main Results}
\label{sec:main_results}

In the 16-shot PENGI evaluation, DF-PALM achieved the highest cross-benchmark mean accuracy (80.69\%) and Macro-F1 (80.32\%) among the evaluated methods. Table~\ref{table:main_results} reports the complete Accuracy results; Table~\ref{table:main_f1} in Appendix~\ref{app:main_f1_results} reports the complete Macro-F1 results. ProKeR achieved 79.07\% accuracy and 78.30\% Macro-F1, whereas FS-AVG achieved 76.06\% and 75.41\%, respectively. DF-PALM led both metrics on seven benchmarks spanning speech emotion, environmental and action sounds, music, surveillance events, and acoustic scenes.

DF-PALM increased cross-benchmark mean accuracy and Macro-F1 over PALM by 4.11 and 4.30 percentage points, respectively (Table~\ref{table:statistical_comparison}). Both 95\% bootstrap percentile intervals excluded zero, and the exact two-sided Wilcoxon signed-rank tests remained significant after Holm correction (both \(p_{\mathrm{adj}}=1.95\times10^{-3}\)).

On the full 11-benchmark suite, DF-PALM increased mean Macro-F1 over ProKeR by \(+2.02\) points (95\% CI, \([0.74,\,3.35]\); \(p_{\mathrm{adj}}=0.037\)), a significant difference after Holm correction. The mean Accuracy difference was \(+1.62\) points (95\% CI, \([0.43,\,2.84]\); \(p_{\mathrm{adj}}=0.067\)). Excluding ESC50-Actions yielded a Macro-F1 gain of 2.17 percentage points over ProKeR (\(p_{\mathrm{adj}}=0.0547\); Appendix~\ref{app:statistics}).

Against FS-AVG, DF-PALM improved Accuracy by \(+4.64\) points (95\% CI, \([1.49,\,8.43]\); \(p_{\mathrm{adj}}=0.037\)) and Macro-F1 by \(+4.91\) points (95\% CI, \([1.63,\,8.85]\); \(p_{\mathrm{adj}}=0.037\)). At the benchmark level, DF-PALM won 8 of 11 paired Accuracy comparisons and 9 of 11 paired Macro-F1 comparisons against ProKeR, and 9 of 11 comparisons on both metrics against FS-AVG. Appendix~\ref{app:statistics} specifies the paired-testing protocol and benchmark-level inputs.

Across all 11 leave-one-benchmark-out omissions, DF-PALM improved mean Accuracy over PALM by 3.75--4.40 percentage points and mean Macro-F1 by 3.87--4.62 points. Excluding ESC50-Actions yielded gains of 4.28 and 4.48 points, respectively (Table~\ref{table:leave_one_out}). Appendix~\ref{app:statistics} reports paired statistics for this subset-omission analysis.

Table~\ref{table:capacity_controls} evaluates DF-PALM against a linear probe, a static output-space MLP, and an encoder-internal LoRA control. All methods were trained and evaluated under the same training settings. DF-PALM exceeded Static-PALM+MLP and PALM+Audio-LoRA by 1.22 and 1.42 accuracy points, and 1.21 and 1.39 Macro-F1 points, respectively. DF-PALM's gains over Static-PALM+MLP and PALM+Audio-LoRA were significant for both metrics after Holm correction. These comparisons support the combination of instance conditioning and class-guided audio refinement over the evaluated static output-space and encoder-internal adaptation controls.

PALM+ASPL$^{*}$ extends the encoder-internal controls with stage-wise audio modulation~\cite{cho2026aspl}. DF-PALM exceeded PALM+ASPL$^{*}$ by 5.12 percentage points in accuracy and 5.24 points in Macro-F1 under the PALM training configuration.

\begin{figure*}[!t]
\centering
\includegraphics[width=0.85\linewidth]{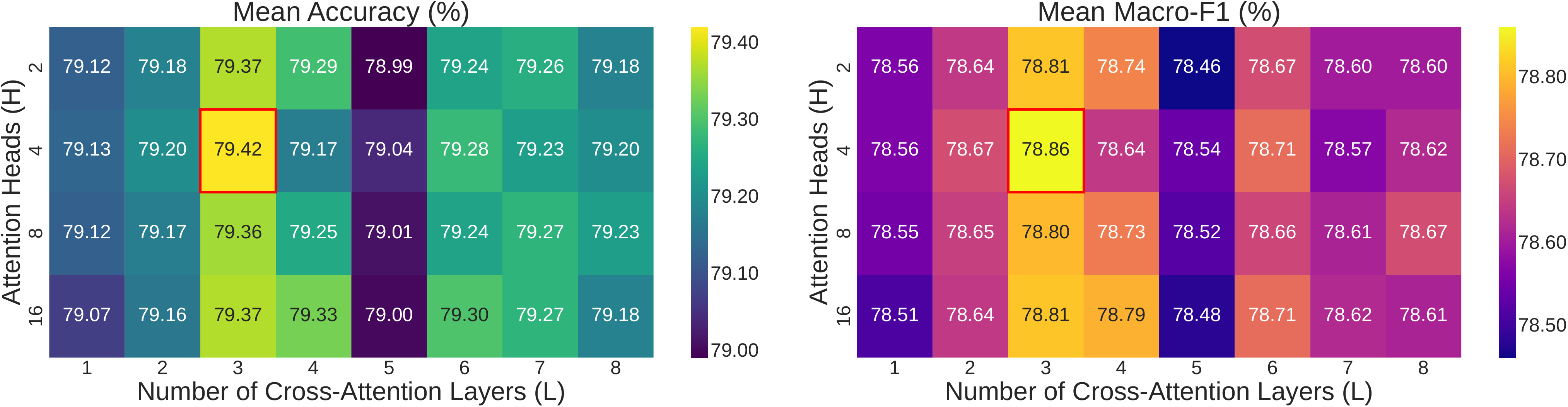}
\caption{Cross-benchmark internal-validation performance across cross-attention depth \(L\) and head count \(H\) in the 16-shot setting. Scores summarize four-fold support-set validation, aggregated across dataset folds and seeds with equal benchmark weighting. Configuration selection is performed independently within each support set. The left and right panels show mean accuracy and Macro-F1, respectively. Red outlines mark \(L=3\), \(H=4\), selected in every independent run.}
\label{fig:hyperparams}
\end{figure*}

\subsection{Cross-Backbone Transfer}
\label{sec:external_backbones}

At 16 shots, DF-PALM improved mean Accuracy and Macro-F1 over PALM across the same 11 benchmarks on all three evaluated audio--language backbones. On LAION-CLAP~\cite{laionclap2023}, DF-PALM increased the mean Accuracy and Macro-F1 by 1.37 and 1.35 percentage points, respectively (Table~\ref{table:external_backbones}). The corresponding gains on OpenFLAM~\cite{flam2025} were 0.90 and 0.74 points. All four paired comparisons remained significant after joint Holm correction. Appendix~\ref{app:external_backbones} reports the benchmark-level differences.

\begin{table}[!t]
\centering
\small
\setlength{\tabcolsep}{2pt}
\caption{Cross-Backbone Comparison at 16 Shots. Scores are unweighted means across 11 benchmarks (\%). Differences and 95\% bootstrap intervals are percentage points. Holm correction covers all four tests.}
\label{table:external_backbones}
\begin{tabular}{@{}l r r r c r@{}}
\toprule
\textbf{Metric} & \textbf{PALM} & \textbf{DF-PALM} & \textbf{$\Delta$} & \textbf{95\% CI} & \textbf{$p_{\mathrm{adj}}$} \\
\midrule
\multicolumn{6}{l}{\textbf{LAION-CLAP}} \\
Accuracy & 75.50 & \textbf{76.87} & +1.37 & [0.71, 1.98] & 0.0098 \\
Macro-F1 & 75.13 & \textbf{76.48} & +1.35 & [0.82, 1.87] & 0.0088 \\
\midrule
\multicolumn{6}{l}{\textbf{OpenFLAM}} \\
Accuracy & 76.94 & \textbf{77.83} & +0.90 & [0.55, 1.31] & 0.0078 \\
Macro-F1 & 76.88 & \textbf{77.62} & +0.74 & [0.43, 1.08] & 0.0098 \\
\bottomrule
\end{tabular}
\end{table}

\begin{table}[!t]
\centering
\small
\caption{Cross-Benchmark Component Ablations and Fusion Control (\%).}
\label{table:ablation}
\begin{tabular}{lcc}
\toprule
\textbf{Method Variant} & \textbf{Mean Accuracy} & \textbf{Macro-F1} \\
\midrule
PALM (Baseline) & 76.58 & 76.02 \\
Dynamic Context only & 79.27 & 78.72 \\
Asymmetric Fusion only & 78.89 & 78.46 \\
Joint Attention & 78.31 & 76.83 \\
\textbf{DF-PALM (Full)} & \textbf{80.69} & \textbf{80.32} \\
\bottomrule
\end{tabular}
\end{table}

\subsection{Performance across Shot Settings}
\label{sec:few_shot}

Across \(K\in\{4,8,16,32\}\) training examples per class, DF-PALM achieved the highest cross-benchmark mean accuracy and Macro-F1 among the four prompt-learning methods (Fig.~\ref{fig:few_shot}). DF-PALM retained each run's 16-shot-selected configuration across shot settings; \(K\) denotes the number of examples per class used for final training. Its paired gains over PALM ranged from 0.63 to 4.11 percentage points in accuracy and from 0.72 to 4.30 points in Macro-F1. All 95\% bootstrap percentile intervals excluded zero. Benchmark-level results for the additional settings are reported in Appendix~\ref{app:shot_results}.

\begin{table}[!t]
\centering
\small
\setlength{\tabcolsep}{2.5pt}
\caption{End-to-End Inference Resource Profile at \(C=50\) and a Batch Size of One on an NVIDIA A40 GPU.}
\label{table:e2e_resources}
\begin{tabular}{lcccc}
\toprule
\textbf{Method} & \textbf{Params (M)} & \textbf{FLOPs} & \textbf{Latency (ms)} & \textbf{Memory} \\
 & \textbf{Total} & \textbf{(G)} & \textbf{Mean \(\boldsymbol{\pm}\) SD} & \textbf{(MB)} \\
\midrule
PALM &  98.216 & 165.348 & \(25.57 \pm 2.60\) & 448.26 \\
DF-PALM & 110.950 & 165.990 & \(28.00 \pm 2.66\) & 497.40 \\
\bottomrule
\end{tabular}
\end{table}

\subsection{Ablation Studies}
\label{sec:ablation}

Both components improved on PALM, and their combination achieved the highest mean scores. Relative to PALM, Dynamic Context alone increased cross-benchmark mean accuracy from 76.58\% to 79.27\% and Macro-F1 from 76.02\% to 78.72\% (Table~\ref{table:ablation}). Asymmetric Cross-Modal Fusion alone achieved 78.89\% and 78.46\%, respectively. Relative to Dynamic Context alone, the full model increased mean accuracy and Macro-F1 by 1.42 and 1.60 percentage points, to 80.69\% and 80.32\%, respectively. The corresponding 95\% bootstrap percentile intervals were \([0.64,\,2.34]\) and \([0.70,\,2.70]\). Both paired tests remained significant after Holm correction (both \(p_{\mathrm{adj}}=3.91\times10^{-3}\)).

Relative to Dynamic Context alone, the full model improved Accuracy on 9 of 11 benchmarks and Macro-F1 on 10 of 11 benchmarks. These results support combining instance-conditioned class representations with class-guided audio refinement, which provides gains beyond conditioning alone.

The joint self-attention control (Table~\ref{table:ablation}) retained the dynamic context generator and matched the fusion depth and head count. It applied self-attention to the audio and class tokens, updating both representations. DF-PALM exceeded this control by 2.38 percentage points in mean Accuracy and 3.48 points in Macro-F1. The paired 95\% intervals were \([-0.18,\,5.80]\) and \([0.46,\,7.19]\), with Holm-adjusted \(p\) values of 0.278 and 0.037, respectively. Holm correction supported the Macro-F1 gain. Appendix~\ref{app:statistics} reports the benchmark-level inputs.

\subsection{Hyperparameter Selection and Sensitivity}

In the main 16-shot setting, we selected the cross-attention depth \(L\) and number of heads \(H\) independently within each matched support set. The mean validation Accuracy across the four internal folds determined each selection; all selections yielded \(L=3,H=4\). Cross-benchmark aggregation of validation scores summarizes configuration sensitivity. Across the evaluated configurations, mean validation Accuracy and Macro-F1 varied by 0.43 and 0.40 percentage points, respectively (Fig.~\ref{fig:hyperparams}). Validation performance therefore varied little across the evaluated depth and head configurations. The full candidate grid and retraining procedure are given in Appendix~\ref{app:implementation}.

\begin{figure}[!t]
\centering
\includegraphics[width=0.70\linewidth]{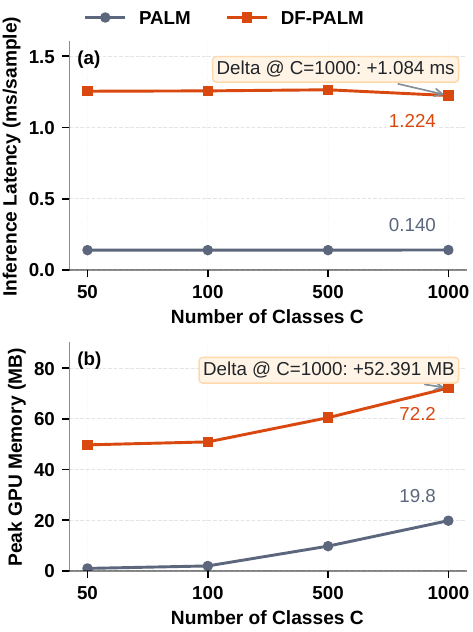}
\caption{Module-level scaling on an NVIDIA RTX 4090 GPU: (a) median latency and (b) peak allocated memory. The frozen PENGI backbone was excluded.}
\label{fig:scalability}
\end{figure}

\subsection{Computational Cost and Scalability}
\label{sec:scalability}

We measured end-to-end inference cost at \(C=50\) and module-level scaling over \(C\in\{50,100,500,1000\}\). Measurement scopes and parameter counts are given in Appendix~\ref{app:profiling}.

At \(C=50\), DF-PALM required \(28.00\pm2.66\)~ms per sample, compared with \(25.57\pm2.60\)~ms for PALM (Table~\ref{table:e2e_resources}). Relative to PALM, operator-counted FLOPs increased by 0.642~GFLOPs (0.39\%), latency by 2.44~ms (approximately 9.5\%), and peak allocated memory by 49.14~MB (10.96\%).

With the frozen backbone excluded, module-level latency changed little across the evaluated class counts (Fig.~\ref{fig:scalability}). At \(C=1000\), DF-PALM required 1.224~ms and 72.2~MB per sample, corresponding to differences of 1.084~ms and 52.391~MB relative to PALM. The complete measurement protocols are given in Appendix~\ref{app:profiling}.

\section{Conclusion}

DF-PALM combines audio-conditioned feature-space prompts with cross-attention that refines audio representations against fixed, instance-conditioned class representations. Both pretrained encoders remain frozen, confining trainable adaptation to feature space. Across 11 benchmarks using PENGI in the 16-shot setting, DF-PALM achieved the highest mean accuracy and Macro-F1 among the evaluated methods, exceeding PALM by 4.11 percentage points in mean accuracy. It also led the four evaluated prompt-learning methods on both metrics across all four shot settings. DF-PALM improved the cross-benchmark means of both metrics over PALM on all three evaluated audio--language backbones.

These results support instance-conditioned feature-space adaptation as an effective approach to few-shot audio classification. At \(C=50\), DF-PALM required 28.00~ms per sample on an NVIDIA A40 GPU, with a 0.39\% increase in operator-counted FLOPs relative to PALM. Module-level latency changed little across the evaluated class counts.

\appendices
\section{Dataset Specifications and Text Inputs}
\label{app:datasets}

The evaluation uses Beijing-Opera~\cite{beijing}, CREMA-D~\cite{crema-d}, ESC50-Actions and ESC-50~\cite{esc50}, GT-Music-Genre~\cite{gt-music}, NS-Instruments~\cite{ns-instru}, RAVDESS~\cite{ravdess}, SESA~\cite{sesa}, TUT2017~\cite{tut}, UrbanSound8K~\cite{urbansound}, and VocalSound~\cite{vocalsound}. Table~\ref{table:app_dataset_overview} specifies class counts and evaluation splits. Fold-based benchmarks draw supports from the non-test folds; other benchmarks use the baseline's archived training and test partitions. Support and test samples are disjoint. ESC50-Actions is the ten-class human-sound subset of ESC-50 used in PALM~\cite{hanif2024palm}. Both benchmarks use the baseline split files without modification.

Class-name strings are loaded unchanged from the \texttt{classname} field and sorted alphabetically. PALM and DF-PALM encode these strings directly as \(p_i\); CoOp and CoCoOp prepend learned context tokens. Zero-Shot and ProKeR use the fixed template specified in the baseline descriptions, whereas FS-AVG uses no text representations.

\begin{table*}[t]
\centering
\small
\setlength{\tabcolsep}{8pt}
\caption{Characteristics and evaluation split structure of the 11 benchmarks. Year denotes the publication year of the benchmark reference cited in this paper; CV denotes cross-validation.}
\label{table:app_dataset_overview}
\begin{tabular}{lclcl}
\toprule
\textbf{Benchmark} &
\textbf{Year} &
\textbf{Domain/task} &
\textbf{Classes} &
\textbf{Split protocol} \\
\midrule

\multicolumn{5}{c}{\textbf{Instrument and Music Classification}} \\
\midrule
Beijing-Opera  & 2014 & Instrument recognition         & 4  & 5-fold CV \\
NS-Instruments & 2017 & Instrument recognition         & 10 & Archived train/test \\
GT-Music-Genre & 2012 & Music genre classification     & 10 & Archived train/test \\

\midrule
\multicolumn{5}{c}{\textbf{Environmental, Urban, and Event Audio}} \\
\midrule
ESC-50          & 2015 & Environmental sound classification & 50 & 5-fold CV \\
ESC50-Actions   & 2015 & Human sound-event classification   & 10 & 5-fold CV \\
UrbanSound8K    & 2014 & Urban sound classification         & 10 & 10-fold CV \\
TUT2017         & 2017 & Acoustic scene classification      & 15 & 4-fold CV \\
SESA            & 2019 & Surveillance sound-event classification & 4 & Archived train/test \\

\midrule
\multicolumn{5}{c}{\textbf{Speech Emotion and Human Vocalizations}} \\
\midrule
CREMA-D    & 2014 & Speech emotion recognition       & 6 & Archived train/test \\
RAVDESS    & 2018 & Speech emotion recognition       & 8 & Archived train/test \\
VocalSound & 2022 & Human vocalization classification & 6 & Archived train/test \\
\bottomrule
\end{tabular}
\end{table*}

\begin{table*}[!tp]
\centering
\caption{Complete Mean Macro-F1 Results (\%) in the Main 16-Shot Setting. Means average folds within each seed and then three matched seeds. AVERAGE weights benchmarks equally; bold marks the highest mean in each row.}
\label{table:main_f1}
{\small
\setlength{\tabcolsep}{8pt}
\begin{tabular}{@{}lrrrrrrc@{}}
\toprule
\multicolumn{1}{@{}l}{\textbf{Benchmark}} & \textbf{Zero-Shot} & \textbf{CoOp} & \textbf{CoCoOp} & \textbf{FS-AVG} & \textbf{ProKeR} & \textbf{PALM} & \multicolumn{1}{c@{}}{\textbf{DF-PALM}} \\
\midrule
Beijing-Opera & 13.21 & 95.21 & 97.63 & \textbf{99.85} & 95.98 & 95.23 & 99.56 \\
CREMA-D & 17.20 & 25.81 & 26.00 & 29.81 & 31.16 & 27.83 & \textbf{35.61} \\
ESC50-Actions & 61.38 & 95.13 & 96.57 & 98.91 & 98.46 & 96.54 & \textbf{99.00} \\
ESC-50 & 44.78 & 93.68 & 94.10 & 96.92 & 96.76 & 95.83 & \textbf{97.10} \\
GT-Music-Genre & 28.07 & 70.93 & 74.74 & 79.55 & 78.15 & 80.20 & \textbf{81.25} \\
NS-Instruments & 26.00 & 59.25 & 61.52 & 66.79 & \textbf{73.16} & 64.38 & 72.88 \\
RAVDESS & 7.71 & 30.43 & 38.07 & 31.83 & 43.52 & 45.46 & \textbf{49.49} \\
SESA & 68.27 & 88.81 & 86.98 & 89.33 & 92.71 & 89.86 & \textbf{94.16} \\
TUT2017 & 17.95 & 64.40 & 73.06 & 67.79 & 79.99 & 78.95 & \textbf{84.10} \\
UrbanSound8K & 51.44 & 75.23 & 76.26 & \textbf{86.99} & 86.53 & 81.10 & 86.67 \\
VocalSound & 38.34 & 70.13 & 77.55 & 81.76 & \textbf{84.87} & 80.86 & 83.68 \\
\specialrule{\lightrulewidth}{0pt}{0pt}
\textbf{AVERAGE} & 34.03 & 69.91 & 72.95 & 75.41 & 78.30 & 76.02 & \textbf{80.32} \\
\bottomrule
\end{tabular}
}
\end{table*}

\section{Implementation and Baseline Adaptation}
\label{app:implementation}

\subsection{Learned Adaptation}
DF-PALM uses a bottleneck reduction ratio of 16 and attention dropout of 0.1. Context vectors use zero-mean normal initialization with standard deviation 0.02; mixing parameters are initialized uniformly on $[0,1)$ before the sigmoid. CoOp and CoCoOp use 16 context tokens. Training and evaluation batch sizes are 16 and 128. Learned methods use cross-entropy and SGD with initial learning rate 0.05, momentum 0.9, and no weight decay. Evaluation uses the epoch-50 model without checkpoint selection and matched seeds 0, 1, and 2.

The linear probe trains a linear classifier on frozen PENGI audio features. Static-PALM+MLP retains PALM's input-independent class representations and uses a three-layer residual audio MLP with approximately 12.60 million trainable parameters. PALM+Audio-LoRA retains the static class representations and adapts the 12 HTSAT~\cite{chen2022htsat} audio-encoder QKV projections with rank-$r=8$ factors, scaling $\alpha=16$, and zero dropout. The 141,312 LoRA parameters are optimized together with the PALM prompt parameters; base encoder weights remain frozen. Inference evaluates the low-rank updates alongside the frozen QKV projections, adding audio-encoder matrix operations without a cross-modal fusion block.

PALM+ASPL$^{*}$ combines static PALM class representations with affine modulation after log-mel extraction, patch embedding, and the first HTSAT stage, including downsampling. The modulation widths are 64, 96, and 192, adding 704 trainable parameters. These parameters and the PALM prompts are optimized jointly, while pretrained encoder weights remain frozen. Table~\ref{table:capacity_controls} reports this implementation under the PALM training configuration.

\subsection{Support-Set Configuration Selection}
For each benchmark, outer fold where applicable, and seed, DF-PALM selects its fusion configuration within the corresponding 16-shot support set. Four internal folds fit each candidate on 12 examples per class and validate on four. Selection maximizes the mean validation Accuracy across the four internal folds for that run, without pooling scores across outer folds, seeds, or benchmarks. The outer test samples are excluded from both candidate fitting and validation. The procedure uses no labels beyond that run's 16-shot support set.

The search covers eight depths and four head counts with the remaining settings fixed. Each independently selected configuration is retrained on all 16 examples per class and evaluated at epoch 50. All selections yielded \(L=3,H=4\). The corresponding configuration is retained for the 4-, 8-, and 32-shot runs. CoOp, CoCoOp, and PALM use their reported hyperparameters without additional tuning. ProKeR selects its kernel configuration using a class-stratified 12/4 support split, then refits on all 16 examples per class. In the main 16-shot setting, the methods share the support-label budget while retaining their specified selection procedures.

\subsection{Training-Free Predictors}
ProKeR applies its kernel formulation~\cite{bendou2025proker} to normalized PENGI support embeddings $\mathbf{A}_S$, query embeddings $\mathbf{A}_Q$, and class-text embeddings $\mathbf{W}$. With one-hot support labels $\mathbf{Y}_S$, its kernel, coefficients, and logits are
\begin{align}
&K_\beta(\mathbf{x},\mathbf{y})=\exp[-\beta(1-\mathbf{x}^{\top}\mathbf{y})],\\
&\boldsymbol{\alpha}=(\lambda^{-1}\mathbf{K}_{SS}+\mathbf{I})^{-1}(\mathbf{Y}_S-\mathbf{A}_S\mathbf{W}^{\top}),\\
&\mathbf{F}_Q=\mathbf{A}_Q\mathbf{W}^{\top}+\mathbf{K}_{QS}\boldsymbol{\alpha}.
\end{align}
The candidate grid comprises 20 linearly spaced $\beta$ values in $[0.5,3.0]$ and 10 linearly spaced $\lambda$ values in $[10^{-3},1]$. Validation Accuracy selects the configuration.

FS-AVG~\cite{taylor25_interspeech} averages unnormalized support audio embeddings per class, L2-normalizes the prototypes, and classifies normalized queries by cosine similarity. It uses no text encoder or configuration selection. Both training-free predictors cache audio embeddings; ProKeR encodes fixed class prompts once per benchmark, fold, and seed. DF-PALM computes fixed text representations once per batch without caching across batches.

\begin{table*}[!tp]
\centering
\caption{Leave-One-Benchmark-Out Robustness at 16 Shots. DF-PALM scores (\%) are unweighted means over the ten retained benchmarks. $\Delta$ denotes DF-PALM minus the indicated reference in percentage points.}
\label{table:leave_one_out}
{\small
\setlength{\tabcolsep}{4pt}
\renewcommand{\arraystretch}{1.08}
\begin{tabular*}{0.90\textwidth}{@{\extracolsep{\fill}}lrrrrrrrr@{}}
\toprule
\multirow{2}{*}{\textbf{Omitted Benchmark}} & \multicolumn{4}{c}{\textbf{Accuracy}} & \multicolumn{4}{c}{\textbf{Macro-F1}} \\
\cmidrule(lr){2-5}\cmidrule(lr){6-9}
 & \textbf{DF-PALM} & $\boldsymbol{\Delta_{\mathrm{PALM}}}$ & $\boldsymbol{\Delta_{\mathrm{ProKeR}}}$ & $\boldsymbol{\Delta_{\mathrm{FS\text{-}AVG}}}$ & \textbf{DF-PALM} & $\boldsymbol{\Delta_{\mathrm{PALM}}}$ & $\boldsymbol{\Delta_{\mathrm{ProKeR}}}$ & $\boldsymbol{\Delta_{\mathrm{FS\text{-}AVG}}}$ \\
\midrule
Beijing-Opera & 78.80 & +4.09 & +1.41 & +5.13 & 78.39 & +4.29 & +1.86 & +5.43 \\
CREMA-D & 84.55 & +3.77 & +1.41 & +4.56 & 84.79 & +3.95 & +1.77 & +4.82 \\
ESC50-Actions & 78.86 & +4.28 & +1.73 & +5.09 & 78.45 & +4.48 & +2.17 & +5.39 \\
ESC-50 & 79.04 & +4.39 & +1.75 & +5.08 & 78.64 & +4.60 & +2.19 & +5.38 \\
GT-Music-Genre & 80.64 & +4.40 & +1.53 & +4.90 & 80.22 & +4.62 & +1.91 & +5.23 \\
NS-Instruments & 81.61 & +3.75 & +1.88 & +4.49 & 81.06 & +3.87 & +2.25 & +4.79 \\
RAVDESS & 83.86 & +4.21 & +1.29 & +3.48 & 83.40 & +4.32 & +1.62 & +3.63 \\
SESA & 79.36 & +4.07 & +1.69 & +4.75 & 78.93 & +4.29 & +2.07 & +4.91 \\
TUT2017 & 80.32 & +3.99 & +1.42 & +3.45 & 79.94 & +4.21 & +1.81 & +3.76 \\
UrbanSound8K & 80.16 & +4.00 & +1.80 & +5.14 & 79.68 & +4.17 & +2.21 & +5.43 \\
VocalSound & 80.38 & +4.22 & +1.90 & +4.92 & 79.98 & +4.44 & +2.34 & +5.20 \\
\bottomrule
\end{tabular*}
}
\end{table*}

\begin{table}[!tp]
\centering
\caption{Benchmark-Level Component and Fusion Controls at 16 Shots. Values are mean Accuracy and Macro-F1 (\%). DF-PALM results appear in Table~\ref{table:main_results} and Table~\ref{table:main_f1}.}
\label{table:app_controls}
{\small
\setlength{\tabcolsep}{6pt}
\begin{tabular}{@{}lrrrr@{}}
\toprule
\multirow{2}{*}{\textbf{Benchmark}} & \multicolumn{2}{c}{\textbf{Dynamic Only}} & \multicolumn{2}{c}{\textbf{Joint Attention}} \\
\cmidrule(lr){2-3}\cmidrule(lr){4-5}
 & Acc. & F1 & Acc. & F1 \\
\midrule
Beijing-Opera & 98.73 & 98.71 & 88.40 & 87.67 \\
CREMA-D & 40.63 & 34.02 & 43.68 & 28.98 \\
ESC50-Actions & 98.42 & 98.40 & 98.75 & 98.74 \\
ESC-50 & 96.68 & 96.60 & 97.12 & 97.02 \\
GT-Music-Genre & 81.17 & 81.10 & 81.83 & 82.07 \\
NS-Instruments & 67.10 & 67.26 & 71.53 & 72.12 \\
RAVDESS & 48.47 & 48.78 & 33.88 & 32.02 \\
SESA & 92.70 & 92.78 & 93.02 & 92.99 \\
TUT2017 & 82.24 & 82.09 & 83.50 & 83.25 \\
UrbanSound8K & 81.93 & 82.35 & 86.06 & 86.73 \\
VocalSound & 83.87 & 83.80 & 83.65 & 83.58 \\
\midrule
\textbf{Mean} & 79.27 & 78.72 & 78.31 & 76.83 \\
\bottomrule
\end{tabular}
}
\end{table}

\section{Complete Macro-F1 Results}
\label{app:main_f1_results}

Table~\ref{table:main_f1} complements the main Accuracy table with mean Macro-F1 scores for all seven methods and all 11 benchmarks.

\begin{table}[!tp]
\centering
\small
\setlength{\tabcolsep}{2.5pt}
\renewcommand{\arraystretch}{1.05}
\caption{Paired Sensitivity Analysis Excluding ESC50-Actions at 16 Shots. $\Delta$ denotes DF-PALM minus reference differences in percentage points over the ten retained benchmarks. Intervals are bootstrap 95\% CIs; Holm correction covers both metrics within each method pair.}
\label{table:overlap_sensitivity}
\begin{tabular}{@{}llcr@{}}
\toprule
\textbf{Reference} & \textbf{Metric} & $\boldsymbol{\Delta}$ \textbf{[95\% CI]} & $\boldsymbol{p_{\mathrm{adj}}}$ \\
\midrule
PALM & Accuracy & \(+4.28\ [2.97,\,5.61]\) & \(0.0039\) \\
 & Macro-F1 & \(+4.48\ [3.08,\,5.93]\) & \(0.0039\) \\
\midrule
ProKeR & Accuracy & \(+1.73\ [0.43,\,3.05]\) & \(0.0840\) \\
 & Macro-F1 & \(+2.17\ [0.77,\,3.58]\) & \(0.0547\) \\
\midrule
FS-AVG & Accuracy & \(+5.09\ [1.72,\,9.12]\) & \(0.0391\) \\
 & Macro-F1 & \(+5.39\ [1.87,\,9.53]\) & \(0.0391\) \\
\bottomrule
\end{tabular}
\end{table}

\section{Paired Statistics and Component Controls}
\label{app:statistics}

Metrics are averaged over dataset folds within each matched seed, then over seeds within each benchmark. For full-suite comparisons, each benchmark forms one statistical unit; cross-benchmark means assign equal weight to all 11 benchmarks. Confidence intervals use 200,000 bootstrap resamples of the paired benchmark differences. Exact two-sided Wilcoxon signed-rank tests remove zero differences and average tied ranks. PENGI comparisons apply Holm correction to Accuracy and Macro-F1 within each method pair. Table~\ref{table:main_results} and Tables~\ref{table:main_f1} and~\ref{table:app_controls} report benchmark means; calculations use unrounded values.

The full 11-benchmark suite is retained for comparison with prior work. A separate sensitivity analysis excludes ESC50-Actions to assess the effect of its overlap with ESC-50. The same paired-testing procedure is applied to the ten retained benchmarks (Table~\ref{table:overlap_sensitivity}).

The controls in Table~\ref{table:app_controls} compare audio refinement and alternative fusion architectures. Dynamic Context Only retains the audio-conditioned class representations but scores them against the unrefined audio feature. Joint Attention retains the context generator and matches the fusion depth and head count, using self-attention over the audio and class tokens to update both representations. DF-PALM instead holds the conditioned class matrix fixed across layers and updates the audio state, as defined in Section~\ref{sec:fusion}. These controls relate the component means in Table~\ref{table:ablation} to their benchmark-level results.

\begin{table*}[!tp]
\centering
\caption{Benchmark-Level DF-PALM Minus PALM Differences at Additional Shot Settings. Values are percentage points after fold and matched-seed aggregation; the mean weights benchmarks equally.}
\label{table:app_shot_differences}
{\small
\setlength{\tabcolsep}{5.2pt}
\renewcommand{\arraystretch}{1.08}
\begin{tabular}{@{}lrrrrrr@{}}
\toprule
\multirow{2}{*}{\textbf{Benchmark}} & \multicolumn{3}{c}{\textbf{Accuracy}} & \multicolumn{3}{c}{\textbf{Macro-F1}} \\
\cmidrule(lr){2-4}\cmidrule(lr){5-7}
 & \textbf{4-Shot} & \textbf{8-Shot} & \textbf{32-Shot} & \textbf{4-Shot} & \textbf{8-Shot} & \textbf{32-Shot} \\
\midrule
Beijing-Opera & -1.97 & +0.42 & 0.00 & -1.96 & +0.39 & 0.00 \\
CREMA-D & -1.21 & +2.69 & -0.18 & -0.73 & +1.44 & +0.56 \\
ESC50-Actions & +2.08 & +0.67 & +0.42 & +2.36 & +0.68 & +0.43 \\
ESC-50 & +1.03 & +0.93 & +0.35 & +1.07 & +0.91 & +0.36 \\
GT-Music-Genre & +1.83 & -0.67 & +1.17 & +1.58 & -1.44 & +0.98 \\
NS-Instruments & +3.92 & +2.35 & +1.56 & +5.44 & +2.41 & +1.54 \\
RAVDESS & +0.61 & +0.20 & -0.88 & +0.86 & 0.00 & -1.05 \\
SESA & +8.25 & +1.59 & +1.90 & +8.96 & +1.59 & +2.37 \\
TUT2017 & +2.53 & +1.85 & +0.72 & +2.42 & +1.88 & +0.73 \\
UrbanSound8K & +3.09 & +2.91 & +1.14 & +2.90 & +2.79 & +1.21 \\
VocalSound & +2.82 & +1.48 & +0.78 & +3.01 & +1.30 & +0.77 \\
\midrule
\textbf{Mean} & \textbf{+2.09} & \textbf{+1.31} & \textbf{+0.63} & \textbf{+2.35} & \textbf{+1.09} & \textbf{+0.72} \\
\bottomrule
\end{tabular}
}
\end{table*}

\begin{table}[!tp]
\centering
\small
\setlength{\tabcolsep}{2.5pt}
\renewcommand{\arraystretch}{1.08}
\caption{Benchmark-Level DF-PALM Minus PALM Differences on the Additional Backbones at 16 Shots. Values are percentage points; the mean weights all 11 benchmarks equally.}
\label{table:app_external_backbones}
\begin{tabular*}{0.95\columnwidth}{@{\extracolsep{\fill}}lrrrr@{}}
\toprule
\multirow{2}{*}{\textbf{Benchmark}} & \multicolumn{2}{c}{\textbf{LAION-CLAP}} & \multicolumn{2}{c}{\textbf{OpenFLAM}} \\
\cmidrule(lr){2-3}\cmidrule(lr){4-5}
 & \textbf{$\Delta$ Acc.} & \textbf{$\Delta$ F1} & \textbf{$\Delta$ Acc.} & \textbf{$\Delta$ F1} \\
\midrule
Beijing-Opera & +0.71 & +0.75 & -0.003 & -0.001 \\
CREMA-D & -0.94 & +0.06 & +2.53\phantom{0} & +1.34\phantom{0} \\
ESC50-Actions & +0.75 & +0.76 & +0.50\phantom{0} & +0.42\phantom{0} \\
ESC-50 & +1.37 & +1.39 & +0.55\phantom{0} & +0.57\phantom{0} \\
GT-Music-Genre & +0.33 & -0.10 & +0.83\phantom{0} & +0.59\phantom{0} \\
NS-Instruments & +2.63 & +2.17 & +0.76\phantom{0} & +0.72\phantom{0} \\
RAVDESS & +2.78 & +2.51 & +0.34\phantom{0} & -0.03\phantom{0} \\
SESA & +2.54 & +2.61 & +1.59\phantom{0} & +1.98\phantom{0} \\
TUT2017 & +1.34 & +1.27 & +0.85\phantom{0} & +0.72\phantom{0} \\
UrbanSound8K & +2.17 & +2.06 & +0.96\phantom{0} & +0.90\phantom{0} \\
VocalSound & +1.39 & +1.36 & +0.95\phantom{0} & +0.89\phantom{0} \\
\midrule
\textbf{Mean} & \textbf{+1.37} & \textbf{+1.35} & \textbf{+0.90}\phantom{0} & \textbf{+0.74}\phantom{0} \\
\bottomrule
\end{tabular*}
\end{table}

\subsection{Leave-One-Benchmark-Out Robustness}
Table~\ref{table:leave_one_out} tests whether the aggregate advantage depends on a single benchmark. Each omission removes the same benchmark from both methods and recomputes the paired mean over the ten retained benchmarks. The trained models and test predictions remain unchanged, so the analysis isolates sensitivity to benchmark composition. Omitting ESC50-Actions or ESC-50 also examines sensitivity to including either of these related benchmarks, whose class overlap is specified in Appendix~\ref{app:datasets}.

All 11 omissions retained positive mean differences against PALM, ProKeR, and FS-AVG on both metrics. The cross-benchmark ordering therefore persisted regardless of which single benchmark was removed. Excluding ESC-50 retained PALM-relative gains of 4.39 Accuracy and 4.60 Macro-F1 points, complementing the ESC50-Actions omission reported in Section~\ref{sec:main_results}.

Table~\ref{table:overlap_sensitivity} reports the paired statistics after excluding ESC50-Actions. Mean gains remained positive against all three reference methods; both metrics remained significant against PALM and FS-AVG.

\section{External Backbone Transfer}
\label{app:external_backbones}

The additional evaluation tests whether the adaptation design remains effective when the pretrained representation source changes. LAION-CLAP and OpenFLAM supply normalized 512-dimensional audio and text features through the interface defined in Section~\ref{sec:method}. The context generator and fusion layers operate on these features without modifying the pretrained encoders. Thus, the same conditioning and update structure is retained across the three evaluated backbones.

Both methods retain the benchmark partitions, matched 16-shot supports, three seeds, and the same training settings. DF-PALM uses the same $L=3$, $H=4$ configuration on both additional backbones without further search. Table~\ref{table:app_external_backbones} reports within-backbone differences, comparing DF-PALM with PALM on the same pretrained representations.

Paired tests follow Appendix~\ref{app:statistics}, with Holm correction applied jointly to the four backbone--metric comparisons. Bootstrap intervals are marginal intervals for each comparison.

DF-PALM improved Accuracy on ten of the eleven benchmarks with each additional backbone. Macro-F1 improved on ten benchmarks with LAION-CLAP and nine with OpenFLAM. The improvements span multiple task domains, including environmental sounds, instruments, acoustic scenes, and human vocalizations. Together with the PENGI results, this task-level evidence supports the effectiveness of the proposed feature-space adaptation across the three evaluated encoder interfaces.

\section{Benchmark-Level Gains at Additional Shot Settings}
\label{app:shot_results}

Table~\ref{table:app_shot_differences} gives the benchmark-level DF-PALM minus PALM gains at 4, 8, and 32 shots. These comparisons retain the fusion configuration selected at 16 shots, testing the same adaptation structure as the number of fitting examples changes. Table~\ref{table:main_results} and Table~\ref{table:main_f1} provide the 16-shot Accuracy and Macro-F1 results, respectively; ProKeR and FS-AVG are evaluated only at that setting.

At four shots, DF-PALM improved both Accuracy and Macro-F1 over PALM on nine of the eleven benchmarks. These gains cover instrument recognition, music genres, environmental sounds, speech emotion, acoustic scenes, and human vocalizations. Mean differences also remained positive at eight and thirty-two shots.

\section{Resource Measurement Scope and Parameter Counts}
\label{app:profiling}

Training times in Table~\ref{table:comparison} cover initialization, data loading, 50 training epochs, evaluation, and I/O with the selected configuration; configuration search is excluded. Methods share the ESC-50 8-shot partition and one RTX 4090 GPU.

The end-to-end profile in Table~\ref{table:e2e_resources} includes the frozen PENGI encoders and uses a preloaded waveform, batch size one, and default floating-point precision on an NVIDIA A40. Each of three runs uses 10 warm-up and 100 timed forward passes with CUDA synchronization. Latency summarizes run medians; memory is peak allocated GPU memory, and FLOPs include only profiler-counted operators. Fig.~\ref{fig:scalability} measures adaptation modules alone on an RTX 4090 using synchronized CUDA events at the same batch size and iteration counts.

At \(C=50\) and \(d=1024\), shared encoders contain 98,165,207 parameters, and PALM optimizes 51,250 class-context and mixing parameters. DF-PALM adds 132,160 generator and 12,601,344 fusion parameters, totaling 12,784,754 optimizer-updated parameters. Counts include biases and normalization parameters and exclude the unused decoder.

% \section*{Acknowledgments}
% This should be a simple paragraph before the References to thank those individuals and institutions who have supported your work on this article.

% \balance

% \clearpage
\bibliographystyle{IEEEtran}
\bibliography{references}

\end{document}